\documentstyle[art8,aaspp4,flushrt,tighten]{article}

\newcommand{\lea}{\lesssim}
\newcommand{\gap}{\gtrsim}

\newcommand{\vth}{v_{\rm th}}
\newcommand{\rast}{R_\ast}

\newcommand{\sila}{\sin\lambda}
\newcommand{\cola}{\cos\lambda}
\newcommand{\tala}{\tan\lambda}
\newcommand{\rwd}{\, r_{\rm wd}}
\newcommand{\Mwd}{M_{\rm wd}}
\newcommand{\msy}{\, {\rm M}_\odot \, {\rm yr}^{-1}}
\newcommand{\Lsun}{\,{\rm L}_\odot}
\newcommand{\rd}{r_{\rm d}}
\newcommand{\dis}{\displaystyle}

\newcommand{\ul}{\underline}

\newcommand{\cri}{_{\rm c}}
\newcommand{\xx}{x}
\newcommand{\XX}{X}
\newcommand{\FF}{{\tilde F}}
\newcommand{\Flux}{F}

\received{}
\accepted{}
\journalid{}{}
\articleid{}{}

\slugcomment{Submitted to the Astrophysical Journal}

\begin{document}

\title{Dynamics of Line-Driven Winds from Disks in Cataclysmic
Variables\\ I. Solution Topology and Wind Geometry}

\author{Achim Feldmeier}

\affil{Department of Physics \& Astronomy, University of Kentucky,
Lexington, KY 40506-0055, USA, E-mail: {\tt achim@pa.uky.edu}}

\and

\author{Isaac Shlosman}

\affil{Department of Physics \& Astronomy, University of Kentucky,
Lexington, KY 40506-0055, USA, E-mail: {\tt shlosman@pa.uky.edu}}

\begin{abstract}

We analyze the dynamics of 2-D stationary, line-driven winds from
accretion disks in cataclysmic variable stars. The driving force is
that of line radiation pressure, in the formalism developed by Castor,
Abbott, \& Klein for O~stars. Our main assumption is that wind helical
streamlines lie on straight cones.

We find that the Euler equation for the disk wind has two eigenvalues,
the mass loss rate and the flow tilt angle with the disk. Both are
calculated self-consistently. The wind is characterized by two
distinct regions, an {\it outer} wind launched beyond four white dwarf
radii from the rotation axis, and an {\it inner} wind launched within
this radius. The inner wind is very steep, up to $80\arcdeg$ with the
disk plane, while the outer wind has a typical tilt of $60\arcdeg$. In
both cases the ray dispersion is small. We, therefore, confirm the
bi-conical geometry of disk winds as suggested by observations and
kinematical modeling. The wind collimation angle appears to be robust
and depends only on the disk temperature stratification. The flow
critical points lie high above the disk for the inner wind, but close
to the disk photosphere for the outer wind. Comparison with existing
kinematical and dynamical models is provided. Mass loss rates from the
disk as well as wind velocity laws are discussed in a subsequent
paper.
  
\end{abstract}


\section{Introduction}
\label{introduction}

Accretion disks are ubiquitous in astrophysical systems ranging from
new-born stars to compact objects, like white dwarfs, neutron stars
and black holes, both stellar and galactic. Due to their high
temperatures and large surface areas, disks appear to be among the
most luminous objects in the Universe. Strong dissipative processes
which accompany accretion around compact objects can eleviate the
radiation energy density in and above the disk, leading naturally to
radiation-driven winds, similar to winds from hot stars.
Observational signatures of such winds have been unambigously detected
in cataclysmic variables (CVs) (Heap~et~al.~1978;
Krautter~et~al.~1981; Klare~et~al.~1982; C\'ordova \& Mason 1982) and
in active galactic nuclei, hereafter AGNs (Arav, Shlosman, \& Weymann
1997, and refs.~therein), but their understanding proved to be
challenging for theorists. In this and the following paper (Feldmeier,
Shlosman, \& Vitello 1999; hereafter Paper~II) we focus on different
aspects of disk winds in CVs, such as their 2-D geometry, solution
topology, mass loss rates and velocity profiles. AGN disks will be
discussed elsewhere.

Theoretical understanding of winds from accretion disks is hampered by
their intrinsically multi-dimensional character and by the richness of
various physical processes supplementing the basic hydrodynamics of
the flow. A number of different driving mechanisms for disk winds have
been predicted and analyzed, from magnetic torques to X-ray disk
irradiation (i.e., Compton-heated and thermally-driven winds) to
resonance line pressure (e.g., Blandford \& Payne 1982; Begelman,
McKee, \& Shields 1983; C\'ordova \& Mason 1985;
Woods~et~al.~1996). Disks in non-magnetic CVs with high accretion
rates, $\gap 10^{-9}\msy$, have an energy output which peaks in the
(far-)ultraviolet, similarly to O, B and WR~stars.  Their spectra
exhibit features which bear similarity to those found in hot and
massive stars, and which are attributed to winds driven by radiation
pressure in resonance and subordinate lines of abundant chemical
elements, so-called line-driven winds (LDWs).  Observational evidence
in favor of LDWs winds from hot stars and disks includes but is not
limited to the P~Cygni line profiles of C~{\sc iv}, N~{\sc v} and
Si~{\sc iv}, ionization levels, high terminal velocities and their
correlation with the luminosity, and UV line behavior during continuum
eclipse in CVs.

Pioneering works by Lucy \& Solomon (1970), Castor (1974) and Castor,
Abbott, \& Klein (1975; CAK in the following) have shown that O~star
winds result from scattering of radiation in the resonance lines of
abundant elements. Elegantly formulated theory of the LDWs from
O~stars by CAK, Cassinelli (1979), Abbott (1980, 1982), Pauldrach,
Puls, \& Kudritzki (1986) and others (for a textbook account, see
Lamers \& Cassinelli 1999), was successfully applied to individual
objects. Further refinements of this theory by Owocki \& Rybicki
(1984, 1985) and Owocki, Castor, \& Rybicki (1988) addressed the issue
of stability of the flow.

First application of the LDWs to accretion disks (Shlosman, Vitello,
\& Shaviv 1985; Vitello \& Shlosman 1988) emphasized the non-spherical
ionizing continuum and driving force as well as a bi-conical geometry
of the outflow. Under a broad range of conditions disk atmospheres in
CVs and AGNs become dynamically unstable because the line opacity
effectively brings them into a super-Eddington regime. Continuum
photons absorbed by the UV resonance lines and re-emitted
isotropically contribute to the momentum transfer to the wind. This
process can be described as a resonant scattering which conserves the
number of photons throughout the wind and results in terminal wind
velocities of the order of the escape speed at the base of the flow.

The dynamics and radiation field of disk LDWs employed by
Shlosman~et~al.~(1985) and by Vitello \& Shlosman (1988) were
oversimplified. Both were approximated by a 1-D planar model allowing
for divergence of the flow streamlines and geometrical dilution of the
radiation field.  Nineteen resonance lines in the range of
$500-1,600$\,\AA\ were included in the calculation of the
self-consistent radiation force. It was noted that disk LDWs are more
restrictive than stellar winds and their development is strongly
governed by the ionization structure in the wind.

Subsequently, a variety of 2-D kinematical models for disk winds in
CVs supplemented by a 3-D radiation transfer in the Sobolev
approximation were explored (Shlosman \& Vitello 1993; Vitello \&
Shlosman 1993).  Calculations using an alternative Monte Carlo
radiation transfer method, albeit with frozen-in ionization, gave
similar results (Knigge, Woods, \& Drew 1995).  Constrained by
synthetic line profiles and by calculated effects of varying basic
physical parameters, such as accretion and mass loss rates,
temperature of the boundary layer, rotation, and inclination angle,
the available phase space for wind solutions was sharply reduced. Wind
resonant scattering regions exhibiting a strongly bi-conical character
regardless of the assumed velocity and radiation fields were
identified and mapped. This allowed to match the observed line shapes
from a number of CVs and to put forward a number of predictions, which
were verified in high-resolution HST observations (Shlosman, Vitello,
\& Mauche 1996; Mauche~et~al.~1999). Most importantly, rotation was
positively identified as the dominant factor shaping the UV line
profiles in CVs confirming that the disk and not the white dwarf is
the wind source.

The above 1-D dynamical and 2-D kinematical modelings suffered from
uniqueness problems which can be removed only by invoking the 2-D
dynamics.  Recent successful attempts by Proga, Stone, \& Drew (1998;
PSD hereafter) to model the 2-D time-dependent radiation hydrodynamics
of disk LDWs was a major breakthrough in our understanding of this
phenomenon. (The model of Murray \& Chiang 1996 for CV winds, on the
other hand, does not provide specifics for the wind geometry and hence
we avoid discussing it here.) It confirmed basically that kinematical
models of disk winds have sampled the correct parameter range and
provided the scaling laws between different wind characteristics,
e.g., between mass loss rate and accretion luminosity, and delineated
the phase space for possible time-dependent solutions. A number of
empirical relationships were put forward which require a physical
explanation.
 
In this paper we focus on the 2-D geometry of a disk LDW in the
presence of a realistic radiation field in CVs. We analyze solutions
of the wind Euler equation, emphasising differences in the solution
topology with that of CAK stellar winds. In the subsequent Paper~II we
address issues related to the mass loss rates and velocity laws of CV
winds. The possible contribution to wind-driving by magnetic stresses
is ignored here (e.g., Blandford \& Payne 1982; Pudritz \& Norman
1986; Emmering, Blandford, \& Shlosman 1992), as are jet-like outflows
seen in other disk systems (Livio 1997).

This paper is organized as follows. Section~\ref{caktheoryforostars}
reviews the relevant aspects of CAK theory for LDWs from 
O~stars. Section~\ref{diskwindgeometry} addresses the 2-D geometry of
disk LDWs, as well as the radiation field above the CV disk.
Section~\ref{verticalwind} deals with an analytic solution for
vertical winds above an isothermal disk, and Section~\ref{tiltedwinds}
analyzes the solution topology and flow geometry for tilted winds
above a disk with a realistic temperature stratification.
Section~\ref{discussion} compares our results with other models and
observations, and Section~\ref{summary} summarizes our work.

\section{CAK theory for O~stars}
\label{caktheoryforostars}

\subsection{The stellar line force}

The CAK theory for line-driven winds from O~stars forms the basis for
our model of CV winds, and is therefore briefly summarized here. CAK
assume a line-distribution function per unit $\nu$ and $\kappa$, from
UV to IR,
 \begin{equation}
 \label{caklinelist}
 N(\nu,\kappa)={1 \over \nu} \; {1 \over \kappa_0}\;\left({\kappa_0
\over \kappa}\right)^{2-\alpha},
 \end{equation}
 \noindent where $\nu$ is the line frequency and $\kappa$ [${\rm cm^2
\, g^{-1}}$] the mass absorption coefficient normalized to $\kappa_0$,
where the latter refers roughly to the strongest driving line in the
flow (Owocki~et~al.~1988). For the power exponent, $0<\alpha<1$ holds,
where the lower limit corresponds to purely optically thin lines, and
the (unrealistic) upper limit to purely optically thick lines. Puls,
Springmann, \& Lennon~(1999) derive $\alpha=2/3$ from Kramers' formula
applied to resonance lines of hydrogenic ions. Similar values of
$\alpha$ are obtained from detailed NLTE calculations for dense
O~supergiant winds (Pauldrach 1987; Pauldrach~et~al.~1994). On the
other hand, for low-density winds, e.g., of B~stars near-the-main
sequence, $\alpha=1/2$ may be more appropriate (Puls, Springmann, \&
Owocki~1998). We shall, therefore, consider both cases $\alpha=1/2$
and $2/3$ to study the effect of $\alpha$ on the structure of disk
winds.

Using eq.~(1), the CAK force from all lines can be written in a
general way applicable for both geometries (e.g., Owocki \& Puls
1996),
 \begin{equation}
 \label{lineforcegeneral} 
 {\bf g}_{\rm L}=\kappa_0\, \vth \; {\Gamma(\alpha) \over 1-\alpha}
\;{1 \over c^2} \int d\omega \; I_\gamma \; \mathbf{\hat\gamma}
\;\tau_\gamma^{-\alpha},
 \end{equation}
 \noindent by means of the Sobolev approximation (Sobolev 1957).
$\Gamma(\alpha)$ is the complete Gamma function, $d\omega$ is a solid
angle centered about $\mathbf{\hat\gamma}$, and $I_\gamma$ is the
frequency-integrated intensity in this direction. The line optical
depth in direction $\mathbf{\hat\gamma}$ is given by
\begin{equation}
 \label{opticaldepth} 
 \tau_\gamma = {\kappa_0 \,\vth \,\rho \over \mathbf{\hat\gamma} \cdot
\nabla(\mathbf{\hat\gamma}\cdot {\bf v})},
 \end{equation}
 \noindent with gas density $\rho$, and $\mathbf{\hat\gamma}\cdot
\nabla (\mathbf{\hat\gamma}\cdot {\bf v})$ being the gradient along
$\mathbf{\hat\gamma}$ of the velocity component in direction
$\mathbf{\hat\gamma}$. Note that $\kappa_0 \,\vth$ is independent of
the ion thermal speed $\vth$, and so is the line force. Assuming
spherical symmetry, and adopting the `radial streaming' approximation
of CAK, i.e., $\tau_\gamma \equiv \tau_r$,
eq.~(\ref{lineforcegeneral}) simplifies to
 \begin{equation}
 \label{lineforceradstr}
 g_{\rm L}=(\kappa_0 \,\vth)^{1-\alpha} \; {\Gamma(\alpha) \over
1-\alpha}\; {\Flux \over c^2} \; \left({dv/dr\over
\rho}\right)^\alpha,
 \end{equation}
 \noindent with frequency-integrated, radial flux $\Flux$.

\subsection{Stellar Euler equation}

For an isothermal, spherically symmetric stellar wind, the stationary
Euler equation in dimensionless form can be written as,
 \begin{equation}
 \label{eulercakfull}
 \left(1-{A^2 \over W}\right) W' = -1 -{4 A^2 \over U} + E W'^\alpha,
 \end{equation}
 \noindent where, after CAK, we introduced a radial coordinate
$U=-\rast/r$, with $\rast$ being the stellar radius. The sound speed,
$A$, and the flow speed, $V=\sqrt{W}$, are normalized to the
photospheric escape speed from the reduced stellar mass,
$M(1-\Gamma)$, where $\Gamma$ is the Eddington factor. The normalized
wind acceleration is given by $W'=dW/dU=$ $r^2 vv'/GM(1-\Gamma)$. Note
that the gravitational acceleration is normalized to $-1$, whereas CAK
normalize it to $-1/2$. The constant $E$ in (\ref{eulercakfull}) is
given by
 \begin{equation}
 \label{eigenvaluecakdef}
 E={\Gamma(\alpha) \over 1-\alpha}\; \left[{\kappa_0\,\vth \over 4\pi
GM(1-\Gamma)}\right]^{1-\alpha}\; {L/c^2 \over \dot M^\alpha},
 \end{equation}
 \noindent where $L$ is the stellar luminosity and $\dot M$ is the
mass loss rate, $\dot M=4\pi r^2 \rho v$. Global solutions to
(\ref{eulercakfull}) exist only above a certain, critical $E\cri$,
called an eigenvalue of the problem, i.e., below a {\it maximum
allowable} mass loss rate. In the zero-sound speed limit, $A=0$, the
differential equation (\ref{eulercakfull}) separates into an algebraic
equation,
 \begin{equation}
 \label{eulercakzerosoundspeed}
 P\equiv W'+1-EW'^\alpha=0,
 \end{equation}
 \noindent and a trivial differential equation $W'=const$, which leads
to the CAK velocity law, $v=v_\infty \sqrt{1-\rast/r}$, where
$v_\infty$ is the flow terminal velocity. The Euler equation in the
form (7) is particularly simple and its terms have a straightforward
physical meaning, namely inertia, gravity and line force.

\subsection{The stellar wind topology: critical point of the flow}
\label{criticalcaksolution}

\begin{figure}
 \epsscale{1.}
 \plotone{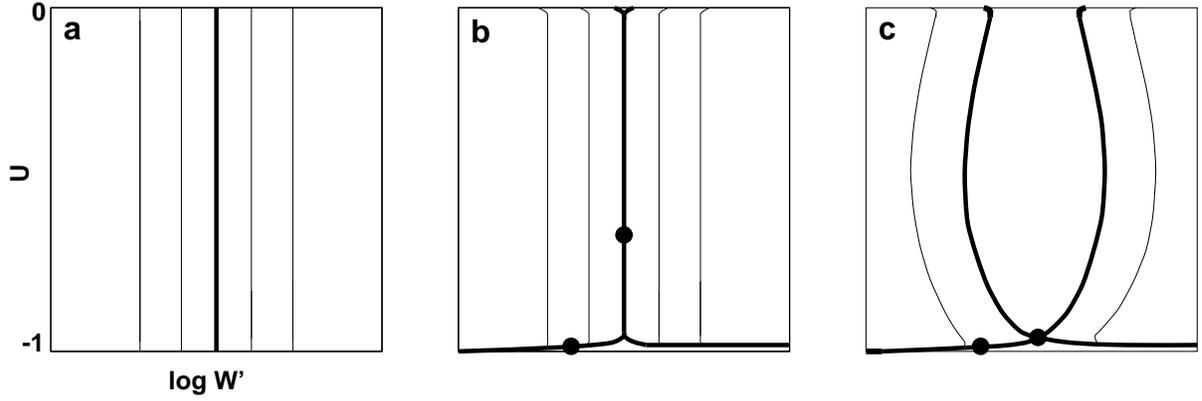}
 \caption{\label{caktopology} Saddle point topology of stellar CAK
winds in the $(W',U)$ plane, for (a) a point star and zero sound
speed; (b) a point star and finite sound speed; (c) an extended star
and finite sound speed.  Filled dots mark the sonic points (at
$U\simeq -1$) and the critical points.}
 \end{figure}

We now consider solutions to eq.~(\ref{eulercakfull}) with finite
$A$. According to CAK, for sufficiently large values of $E$ there are
two solutions in the supersonic regime, $W>A^2$, termed `shallow'
(small $W'$) and `steep' (large $W'$) solutions, whereas in the
subsonic, photospheric regime, $W<A^2$, only the shallow solutions
exist. On the other hand, only the steep solutions reach
infinity. Namely, the term $-4 A^2/U$ (the thermal pressure force due
to geometrical expansion) becomes infinite for $r \rightarrow \infty$,
and must be balanced by $W' \rightarrow \infty$ along the branch of
steep solutions. CAK concluded therefore that the true, unique wind
solution has to switch from the shallow to the steep branch at a
`critical' point (see Fig.~\ref{caktopology}). Of course, a non-zero
pressure term at infinity is unphysical because it requires an
infinite amount of energy in the flow and is purely a result of the
imposed isothermal conditions in the wind.

The subsonic region has an extent of a few percent of the stellar
radius for O~star winds, while the pressure force $-4 A^2/U$ becomes
important only beyond a few $100 \,\rast$. In the intermediate regime,
i.e., almost everywhere, the simplified
eq.~(\ref{eulercakzerosoundspeed}) with $A = 0$ holds to a good
approximation, and therefore $W'= const$. This happens since both
gravity and line force are $\propto r^{-2}$. Solution curves in the
$(W',U)$ plane are essentially straight lines, bending over due to
thermal pressure both at $U=-1$ and 0. As a result, the critical point
is a saddle point in the $(W',U)$ plane (Bjorkman 1995). The usual
mathematical definition of a critical point refers, however, to the
$(W,U)$ plane or a topologically equivalent plane.

For $A=0$, the critical point lies on an extended ridge, and its
position becomes ill-defined (Fig.~\ref{caktopology}a). In this limit,
{\it every} point of the critical solution is a critical point. For
$A>0$, however, CAK find the critical point to lie at $r\cri={3\over
2}\rast$ (Fig.~\ref{caktopology}b). Inclusion of further correction
terms to the line force, especially due to the finite size of the
stellar disk, breaks the $r^{-2}$ dependence of the line force, pushes
the two almost degenerate critical solutions $W'(U)$ apart, and shifts
the critical point towards the sonic point
(Fig.~\ref{caktopology}c). Pauldrach~et~al.~(1986) find then $r\cri
\lea 1.1 \,\rast$.

A critical point is the information barrier for LDWs, and plays a role
similar to the sonic point in thermal winds or nozzle flows (Abbott
1980). How can the pressure mismatch of a shallow solution at $r\gap
300 \,\rast$ be communicated upstream to the critical point at $1.1
\,\rast$? We speculate that it is not really the outer boundary
mismatch which forces the flow through a critical point.  Instead, the
truly distinguishing property of the critical solution should be its
correspondence to the {\it maximum} mass loss rate in the wind. Work
is underway to identify the feedback mechanism between the wind and
the photosphere which drives the wind from any shallow to the unique
critical solution. This issue will be addressed elsewhere. In the
present paper, we {\it assume} that the true disk wind solution is the
one with the maximal allowable mass loss rate. Additional
justification comes from the fact that only the shallow solutions for
the disk wind connect to the disk photosphere. However, their terminal
speeds are much smaller than the white dwarf escape speed, in sharp
contrast to observed CV winds. The solution we are searching for
should therefore switch to the steep branch (with large $v_\infty$) at
some critical point, i.e., should be the solution of maximum mass loss
rate.

The flow critical point (subscript `c') is defined by the singularity
condition, $\partial P/\partial W'\cri=0$ (i.e., merging of a shallow
and a steep solution). Together with the Euler equation, this implies
 \begin{mathletters}
 \begin{eqnarray}
 \label{eigenvaluecakvalue}
 W'\cri&=&{\alpha \over 1-\alpha},\\
 E\cri&=&{1 \over \alpha^\alpha (1-\alpha)^{1-\alpha}}.
 \end{eqnarray}
 \end{mathletters}
 \noindent The eigenvalue $E\cri$ determines the maximum mass loss
rate, and $W'\cri$ determines the terminal speed. They are further
discussed in Paper~II.  Furthermore, from the Euler equation, $P=0$,
also $dP/dU=0$ must hold everywhere. This leads to the regularity
condition, $\partial P/\partial U\cri + W'\cri\; \partial P/\partial
W\cri=0$ (if $W\cri''<\infty$), which determines the position of the
critical point.

\section{Disk wind geometry and radiation field in CVs}
\label{diskwindgeometry}

\subsection{Flow geometry, gravity and centrifugal force}

The central assumption throughout this paper is that the helical
streamline of a fluid parcel in the wind is contained within a {\it
straight} cone. While this is certainly an idealization, and a major
restriction of this model, justifaction comes firstly from the related
kinematical model of Shlosman \& Vitello (1993); and secondly from the
numerical 2-D hydrodynamic simulations of PSD, which showed that the
escaping mass-loss carrying streamlines are well approximated by
straight lines in the $(r,z)$ plane, $r$ and $z$ being cylindrical
coordinates.

\begin{figure}
 \epsscale{1.}
 \plotone{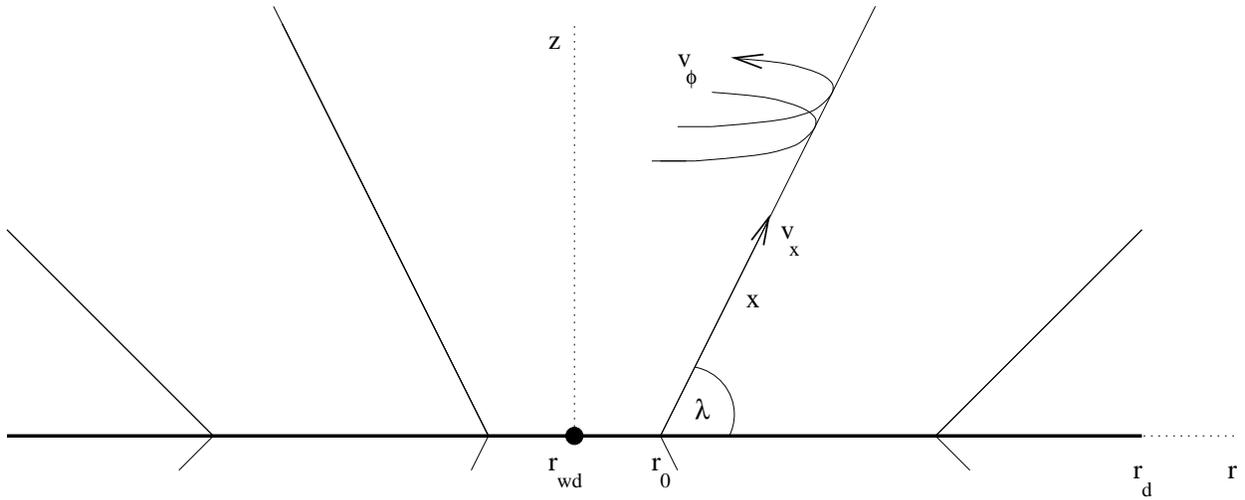}
 \caption{\label{windcoordinates} Adopted flow geometry for a CV disk
wind. The streamlines are helical lines, and are assumed to lie on
straight cones.}
 \end{figure}

We denote the angle between the wind cone and the radial direction
${\bf\hat r}$ in the disk plane by $\lambda$. This angle is calculated
self-consistently using the Euler equation, and is not assumed {\it a
priori}. The footpoint radius of a streamline in the disk is $r_0$,
and $\xx$ is the distance along the cone
(cf.~Fig.~\ref{windcoordinates}).  We search for the solutions of the
Euler equation in the $[\lambda(r_0), \xx]$ plane, for a streamline
starting at arbitrary $r_0$. The $\lambda(r_0)$ dependence leads to
the appearance of a new eigenvalue problem for the disk wind, and
derivation of this function is the focus of the present paper.

Since LDWs are highly supersonic, we neglect the pressure forces, and
furthermore assume that the azimuthal velocity is determined by
angular momentum conservation above the disk, and by Keplerian
rotation in the disk plane. The tilt angle $\lambda$ has to be a
monotonically decreasing function of $r_0$ to avoid streamline
crossing, which would violate the assumption of a pressureless
gas. The only remaining velocity component is $v_\xx$, which points
upwards along the cone. The dynamical problem has therefore been
reduced to solving the Euler equation for $v_\xx$.

In a frame rotating with the angular velocity ${\bf\Omega}$ of a fluid
parcel positioned at radius vector ${\bf r}$, there are three
fictitious forces (e.g., Binney \& Tremaine 1987). The Coriolis
acceleration, $-2\;{\bf\Omega} \times \dot{\bf r}$, has no component
along ${\bf\hat\xx}$, and the same is true for the inertial force of
rotation, $-\, \dot{\bf\Omega} \times {\bf r}$. We introduce the
effective gravity function, which is the component of gravity minus
centrifugal force along the straight line in direction ${\bf\hat\xx}$,
and is given by $-(G\,\Mwd/r_0^2) \, g(\XX,\lambda)$, with $\Mwd$
being the mass of the white dwarf, and
 \begin{equation}
 \label{normalizedgravity} 
 g(\XX,\lambda)= {\XX+\cola \over (1+2\XX \cola+\XX^2)^{3/2}} - {\cola
\over (1+\XX\cola)^3},
 \end{equation}
 \noindent and $\XX\equiv \xx/r_0$. In the following, all lengths
written in capital letters are normalized to the footpoint radius,
$r_0$, in the disk.  Close to the disk, $\XX \ll 1$, and $g\simeq
\XX$, while for $\XX \gg 1$, $g\simeq \XX^{-2}$. For a vertical ray,
$\lambda=90\arcdeg$, $g$ has its maximum at $\XX=1/\sqrt{2}$, while
for a horizontal ray, $\lambda=0$, the maximum is at $\XX=1/2$.

Equation~(\ref{normalizedgravity}) shows an important difference
between stellar and disk winds. The stellar gravity is always
decreasing with distance, while for the disk an effective gravity
`hill' must be overcome before the wind can escape. This effect of
disk LDWs will be discussed in Section~\ref{tiltedwinds}.

\subsection{Radiation field above the disk} 

Next, we evaluate the line force in eq.~(\ref{lineforcegeneral}).
Besides the initial growth of effective gravity with height, the
opacity-weighted flux integral is the central property which
distinguishes disk winds from stellar winds. Pereyra, Kallman, \&
Blondin~(1997) give an analytical approximation for this integral
above an isothermal disk. Unfortunately, an error was introduced with
a change of integration variable, which led to an artificial, linear
dependence of the vertical flux on $z$, even as $z\rightarrow 0$. PSD
solve this integral numerically (cf. Icke 1980), using approximately
2,000 Gaussian integration points.

In the general spirit of the radial streaming approximation of CAK, we
replace the integral in eq.~(\ref{lineforcegeneral}) by
$\bar\tau^{-\alpha} {\bf\Flux}$, thus introducing an {\it equivalent}
optical depth, $\bar\tau$.  We first calculate the
frequency-integrated flux $d {\bf\Flux} (r,z)$ at location $(r,z)$
from a flat ring of radius $q$, radial width $dq$, and isotropic
intensity $I(q,0)$,
 \begin{equation}
 \label{fluxring} 
 d{\bf\Flux}(r,z)=\left(\begin{array}{c} d\Flux_r \\ d\Flux_z
\end{array}\right)=2\pi\, I(q,0)\,q\,dq\, {z \over B^{3/2}}
\left(\begin{array}{c} r\bigl[r^2+z^2-q^2\bigr] \\
z\bigl[r^2+z^2+q^2\bigr] \end{array}\right),
 \end{equation}
 \noindent where
 \begin{equation}
 \label{capitalx} 
 B=\bigl(r^2+z^2-q^2\bigr)^2 + 4z^2q^2.
 \end{equation}
 \noindent For an {\it isothermal} disk with isotropic intensity, we
integrate eq.~(\ref{fluxring}) over $q$, to obtain the disk flux
 \begin{equation}
 \label{fluxiso} 
 {\bf\Flux}(r,z)={\pi \, I \over 2}\; {1 \over \sqrt{B}}
\left.\left(\begin{array}{c} {\dis z \over \dis
r}\bigl[r^2+z^2+q^2\bigr)] \\ -r^2-z^2+q^2\end{array} \right)
\right|_{q=\rwd}^{\rd},
 \end{equation}
 \noindent where $\rd$ is the outer disk radius. For the nonmagnetic
systems considered here, we identify the inner disk radius with
$\rwd$, the white dwarf radius. We do not include contributions to the
radiative flux from the white dwarf and the boundary layer. Generally,
of course, accretion disks are not isothermal. We, therefore, consider
two complementary cases with $T(r_0)\propto r_0^{-1/2}$ (termed
`Newtonian' disk in the following) and $T(r_0)\propto r_0^{-3/4}
\;\left(1-\sqrt{\rwd/r_0}\right)^{1/4}$ (Shakura \& Sunyaev 1973; SHS
hereafter).  Observations show that the brightness temperature
stratification of CV disks is consistent with both these distributions
(Horne \& Stiening 1985; Horne \& Cook 1985; Rutten~et~al.~1993).

For the Newtonian disk, we find
 \begin{equation}
 \label{fluxrsq}
 {\bf\Flux}(r,z)=\pi \, I(r) \, r^2 \,{z \over r^2+z^2} \left.\left
(\begin{array}{l}{\dis 3r^2-z^2-q^2 \over \dis 2r\sqrt{B}} - {\dis r
\over \dis z^2+r^2} \ln \; C \\ \phantom{0}\\ {\dis 3z^2-r^2+q^2 \over
\dis 2z\sqrt{B}} - {\dis z \over \dis z^2+r^2} \ln \; C \end{array}
\right) \right|_{q=\rwd}^{\rd},
 \end{equation}
 \noindent where 
 \begin{equation}
 \label{defy}
 C={\dis \bigl(z^2-r^2\bigr)q^2 +
\bigl(z^2+r^2\bigr)^2 + \bigl(z^2+r^2\bigr) \sqrt{B} \over \dis q^2}.
 \end{equation}

The surface flux above the SHS disk can only be integrated
numerically, from eq.~(\ref{fluxring}). Yet, this has the advantage
that $\bar\tau$ can be introduced for each ring individually. More
specifically, $\bar\tau$ is calculated along the flux direction of a
given ring at the position of the wind parcel. If the flux
(\ref{fluxrsq}) is used instead, $\bar\tau$ is calculated along the
disk flux direction. Typical differences in the resulting value for
the tilt angle $\lambda\cri$ (see below) are $5\arcdeg$ to $10\arcdeg$
for the two approaches. Corrections due to the $\tau^{-\alpha}$
weighting in the azimuthal integral are even smaller.

\begin{figure}
 \epsscale{1.}
 \plotone{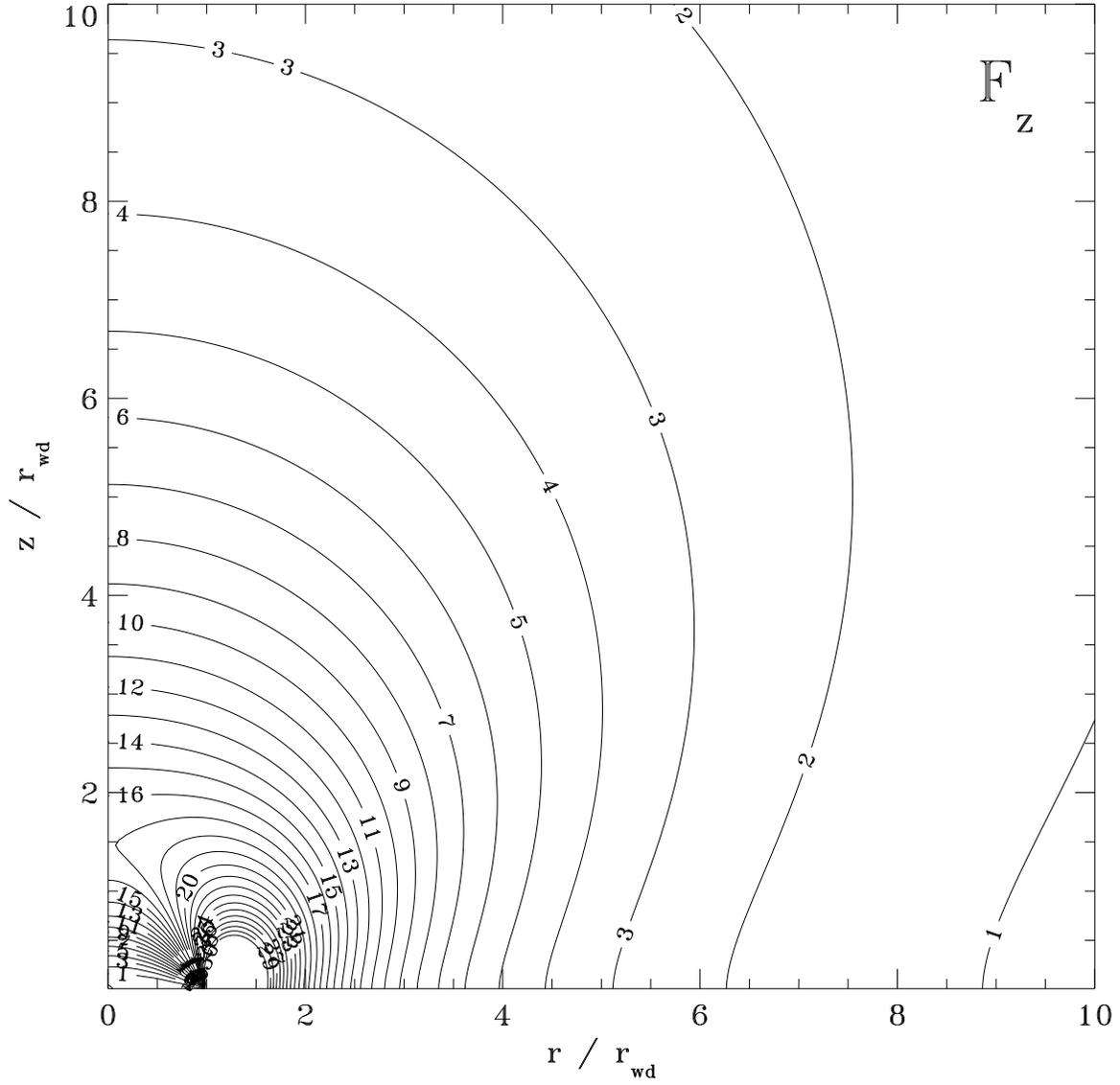}
 \caption{\label{fluxisocontoursz} Isocontours of the
frequency-integrated, vertical flux component $\Flux_z$ above a
Newtonian disk. The disk extends from 1 to $30\rwd$. Normalization is
$I(r_0=5\rwd,z=0)=1$.}
 \end{figure}

\begin{figure}
 \epsscale{1.}
 \plotone{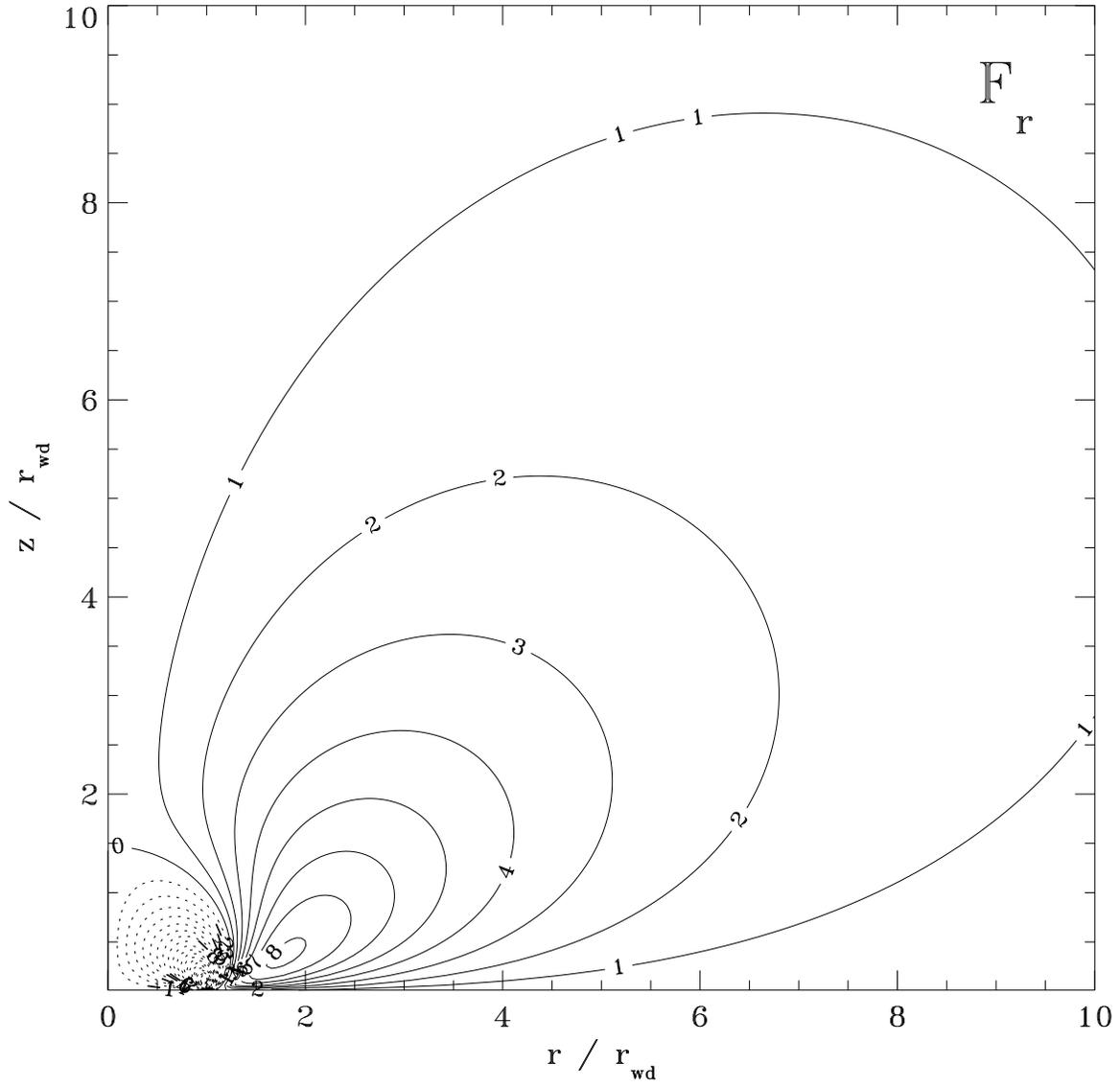}
 \caption{\label{fluxisocontoursr} Same as
Fig.~\ref{fluxisocontoursz}, including normalization, here for the
radial flux component $\Flux_r$. Dotted lines indicate an inward
flux.}
 \end{figure}

Figures~\ref{fluxisocontoursz} and \ref{fluxisocontoursr} show
isocontours for the $z$- and $r$-components of the flux
(\ref{fluxrsq}) above the Newtonian disk. For sufficiently large tilt
angles, the flux along the streamlines has a maximum larger than $\pi
I(r_0,0)$ at some $\XX$. This is due to the increasing visibility of
the inner, hot disk regions. We denote this regime, where the flux has
a maximum, the `panoramic' regime, to be distinguished from the planar
`disk' regime, where $\Flux_z\simeq \pi I$, and the `far field'
regime, where $\Flux\propto \XX^{-2}$.

\begin{figure}
 \epsscale{.7}
 \plotone{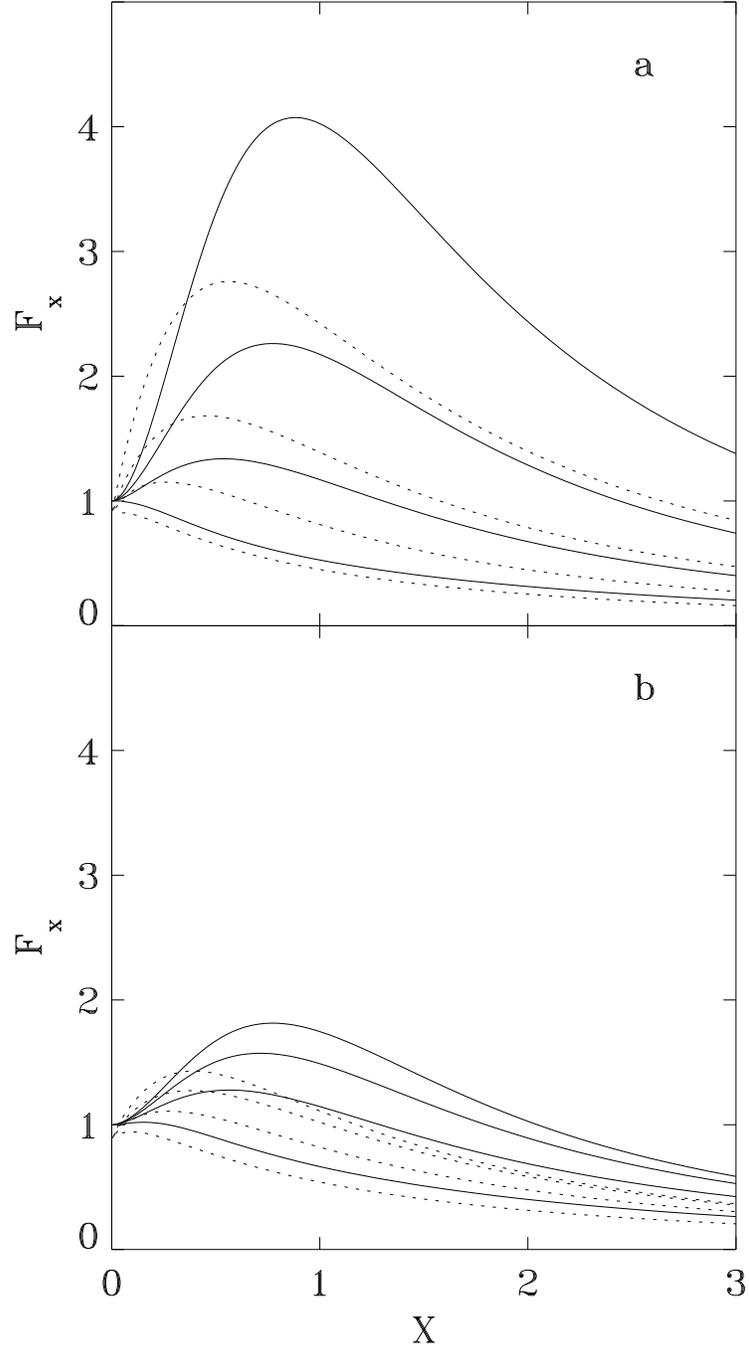}
 \caption{\label{ffunctionplot} Normalized, projected flux
$\FF_\xx(r_0,\XX)$ for SHS (top panel) and Newtonian (bottom panel)
disks, at footpoint radii $r_0=20$, 10, 5, and $2\rwd$ (top to bottom
curves). The tilt angle with the disk plane is $90\arcdeg$ (full
lines) and $60\arcdeg$ (dashed lines).}
 \end{figure}

We now introduce the normalized flux, ${\bf \FF}(r,z)= {\bf \Flux}
(r,z)/\pi I(r_0,0)$, along a streamline, which is independent of disk
luminosity. To quantify the flux increase above the disk,
Fig.~\ref{ffunctionplot} shows the projected, normalized flux
$\FF_\xx= $ ${\bf\hat\xx} \cdot {\bf \FF}$ as function of $\XX$, for
different footpoint radii $r_0$ and for both types of non-isothermal
disks. Note that the initial increase of $\FF_\xx$ with $\XX$ due to
the wind's exposure to the central disk region is rather mild, a
factor of a few only, because the central region has a small area. In
Section~\ref{tiltedwinds} we discuss how the maximum in $\FF_\xx$
controls the base extent of the wind above the disk, as well as the
height of the wind critical point.

In deriving the above fluxes, the intensity was assumed to be
isotropic. Using instead the Eddington limb darkening law,
$I_\theta=\case{2}{5} I_0 \left(1+\case{3}{2} \cos\theta\right)$, with
polar angle $\theta$, the vertical flux in the planar disk regime
above an isothermal disk becomes larger by a factor of 8/7, i.e., limb
darkening should not significantly affect the wind
properties. However, limb darkening can be more important in the UV
spectral regime due to the Wien part of the spectrum, and the
correction factors could become somewhat larger there (Diaz, Wade, \&
Hubeny~1996).

\section{Vertical wind above an isothermal disk}
\label{verticalwind}

As an analytically tractable case, we consider first a vertical (or
cylindrical) wind with $\lambda=90\arcdeg$, or ${\bf\hat\xx}= {\bf\hat
z}$, above an infinite, isothermal disk with a flux ${\bf \Flux}=\pi I
\; {\bf\hat z}$. We again adopt the `radial streaming' approximation
in (\ref{lineforcegeneral}), i.e., $\bar\tau=\tau_z$. Note that
$\tau_z$ has no contributions from either azimuthal velocity
gradients, $\partial v_\phi/\partial r$, or from geometrical expansion
terms, $\propto v_\phi$, the latter describing photon escape along the
tangent to the helical streamline.

The density $\rho$ which enters $\tau$ is replaced by introducing the
mass loss rate from one side of a disk annulus, $d\dot M$. Since the
mass which streams upwards between two cylinders is conserved (using
again coordinate $x$),
 \begin{equation}
 \label{diskmasslossratecyl} 
 d\dot M(r_0)=2\pi\, r_0 \,dr_0\, v_\xx(r_0,\xx)\,\rho(r_0,\xx).
 \end{equation}
 \noindent For simplicity, we apply the zero sound speed limit, $A=0$,
for the rest of this paper, and neglect the force due to electron
scattering because of small $\Gamma$ above the geometrically-thin
disk.  The Euler equation becomes
 \begin{equation} 
 \label{eulerforverticaldiskwind}
 0=P(W',\XX)=W'+g-E\, W'^\alpha,
 \end{equation}
 \noindent where $g={\XX/(1+\XX^2)^{3/2}}$ for $\lambda=90\arcdeg$,
and $W'=2V\,{dV/d\XX}$. Here $d/d\XX=r_0\;d/d\xx$, and $V$ is the flow
speed normalized to the {\it local} escape speed at the footpoint
$r_0$ on the disk. Normalizing instead to the escape speed from the
white dwarf leads to unwanted, explicit appearances of $r_0$ in the
Euler equation. The constant $E(r_0)$ for a streamline starting at
$r_0$ on the disk is defined as (compare with
eq.~\ref{eigenvaluecakdef})
 \begin{equation} 
 \label{eigenvaluediskdef}
 E(r_0)= {\Gamma(\alpha) \over 1-\alpha}\;\left({\kappa_0 \,\vth
\over 2\pi G\,\Mwd}\right)^{1-\alpha} \;{2\pi\, r_0^2\, \Flux_z
(r_0,z=0)/c^2 \over (r_0\; d\dot M(r_0)/dr_0)^\alpha}.
 \end{equation} 

Similarly to the stellar case, eq.~(\ref{eulerforverticaldiskwind})
for the disk wind has solutions only above a critical value $E\cri$,
the eigenvalue of the problem, i.e., below a maximum allowable mass
loss rate. Unlike the point star case, however, $P$ in
eq.~(\ref{eulerforverticaldiskwind}) is a function of $\XX$ even when
$A=0$. As a result, the degeneracy in the position of the critical
point does not exist here, and one has a well-defined critical point,
irrespective of $A$.

There exists an additional difference between the stellar and disk
LDWs. In Fig.~\ref{caktopology}c, the finite cone correction factor
causes the critical point in the stellar wind to move upstream, and,
for vanishing sound speed, the critical point and the sonic point both
are found in the stellar photosphere. For the disk case, however, only
the sonic point falls towards the photosphere, whereas the critical
point stays at finite height. Namely, from the regularity condition
$\partial P/\partial \XX=0$ (since $P$ does not depend on $W$), the
critical point of the disk wind lies at the location of maximal
gravity, at $\XX\cri=1/\sqrt{2}$.

This explains why Vitello \& Shlosman (1988) find no critical point in
the disk regime, $\XX \ll 1$, for a vertical wind with a constant
ionization. The variable wind ionization introduces additional
gradients into the driving force, shifting the critical point towards
the disk photosphere. For the solution discussed here, vertical
ionization gradients are not mandatory.

The conditions $P=0$ and $\partial P/\partial W'=0$ lead to 
 \begin{mathletters}
 \begin{eqnarray}
 \label{eigenvalueverticalwinda}
 W'\cri &=& \;{\alpha \over 1-\alpha} \; g\cri  ,\\
 \label{eigenvalueverticalwindb}
 E\cri &=& {1 \over \alpha^\alpha (1-\alpha)^{1-\alpha}} \;
g\cri^{1-\alpha},
 \end{eqnarray}
 \end{mathletters}
 \noindent where $g\cri=2/(3\sqrt{3})$. This defines the wind solution
of maximum allowable mass loss rate. The effective gravity hill
imposes a `bottleneck' on the flow, i.e., the maximum of $g(\XX)$
defines the minimum, constant eigenvalue $E\cri$, or the maximum
allowable $\dot M$, for the critical solution which extends from the
disk photosphere to arbitrary large $\XX$. Larger values of $E$
correspond to shallow solutions and hence to smaller mass loss
rates. Smaller values of $E$ correspond to stalling wind solutions.
Note that $E\cri$ in eq.~(\ref{eigenvalueverticalwindb}) is
independent of $r_0$, in accordance with
eq.~(\ref{eulerforverticaldiskwind}).

\section{Tilted disk winds}
\label{tiltedwinds}

With all pre-requisites at hand, we can now solve the general
eigenvalue problem for a tilted wind above a non-isothermal disk. The
density $\rho$ in (\ref{opticaldepth}) is replaced by the conserved
mass-loss rate between two wind cones,
 \begin{equation}
 \label{diskmasslossrate} 
 d\dot M(r_0)=2\pi r_0 dr_0 \;(1+\XX\cola) \left[1-{\dis \XX r_0
\,(d\lambda/dr_0) \over \dis \sila}\right]\; \sila\; v_\xx(r_0,\xx)\;
\rho(r_0,\xx).
 \end{equation}
 \noindent The term ($1+\XX\cola$) describes the density drop due to
the increasing radius of the cone, and [$1-\XX r_0\, (d\lambda/dr_0)/
\sin\lambda]$ describes the density drop due to the geometrical
divergence of neighboring cones. The factor $\sila$ stems from the
quenching of the flow at small $\lambda$.

\subsection{Disk Euler equation}

The geometrical expansion term $\nabla\mathbf{\hat\gamma}$ in the
directional derivative $\mathbf{\hat\gamma}\cdot \nabla
(\mathbf{\hat\gamma} \cdot {\bf v})$ has contributions from the
azimuthal curvature of helical streamlines and from the cone
divergence $d\lambda/dr_0$. {\it Close} to the disk, where the mass
loss rate of the wind is established, both contributions are
small. For azimuthal curvature terms, this is shown in Appendix~A.
With regard to cone divergence, the argument is {\it a posteriori},
i.e., we find below that $d\lambda/dr_0$ is small. Two neighboring
wind rays launched at, e.g., $r_0\sim 5\rwd$ intersect at a normalized
distance $\XX_{\rm i} \sim -10$ below the disk.  Generally, $\XX_{\rm
i}$ is larger by a factor of 10 than $\XX\cri$, the distance between
the disk plane and the critical point.  By analogy with spherically
symmetric stellar winds, where $\mathbf{\hat\gamma}\cdot \nabla
(\mathbf{\hat\gamma} \cdot {\bf\hat r} \, v_r)=$ $\mu^2 \, dv_r/dr +
(1-\mu^2)\, v_r/r$, with $\mu=\mathbf{\hat\gamma}\cdot {\bf\hat r}$,
the geometrical expansion term $\propto v_\xx \, \nabla
(\mathbf{\hat\gamma} \cdot {\bf\hat\xx})$ for disk winds should be
$\propto v_\xx/[r_0 (X-X_{\rm i})]$. Whereas the geometrical expansion
term for O~star winds, $(1-\mu^2)\, v_r/r$, is of the same order as
the gradient term, $\mu^2 \, dv_r/dr$, it is much smaller for disk
winds. On the other hand, {\it far} from the disk, the expansion term
may become important. However, we find from solving the Euler equation
that it has only a marginal influence on the terminal wind
velocity. Azimuthal terms for helical streamlines are unimportant far
from the disk, where the wind is essentially radial. We, therefore,
neglect all geometrical expansion terms in the following.  Appendix~A
also shows that gradients in the azimuthal velocity can be neglected
in the line force. Finally, we assume that the gradient of $v_\xx$
points in the ${\bf\hat\xx}$-direction.  This is a reasonable
assumption since the velocity gradients develop roughly in the flux
direction, as is also shown below. The normalized Euler equation for a
conical disk wind, and for vanishing sound speed, is then
 \begin{eqnarray} 
 \label{eulerdiskwind}
 &&0=P(W',\XX)=W'+g-E\,f\,W'^\alpha,
 \end{eqnarray} 
 \noindent with auxiliary function $f$,
 \begin{eqnarray}
 \label{ffunctioneq} 
 f(r_0,\XX)\equiv \left[(1+\XX\cola)\; \Bigl(\sila-\XX r_0\, {d\lambda
\over dr_0}\Bigr)\right]^\alpha \; \int\limits_{\rwd}^{\rd} d\FF\,
\mu^{1+2\alpha}.
 \end{eqnarray}
 \noindent Here, $d\FF=|d{\bf\FF}|$ (see eq.~\ref{fluxring}), and
$\mu$ is the cosine of the angle between $d{\bf\FF}$ and the wind
cone. Again, $W'=2V\,{dV/d\XX}$), where the velocity $V$ is normalized
to the local escape speed; $E(r_0)$ is defined in
eq.~(\ref{eigenvaluediskdef}). Note that the flux integral in
(\ref{ffunctioneq}) introduces a further dependence of $f$ on
$r_0$. Furthermore, due to the weighting with $\mu^{2\alpha}$ in the
integral, the {\it disk} flux vector and the wind cone do not
generally point in the same direction. For disk winds as considered
here, good alignment between radiative flux and wind flow is expected,
however. In cases where such an alignment is not possible, e.g., for
atmospheres irradiated from above, ablation winds at large tilt angle
with the radiative flux were recently suggested (Gayley, Owocki, \&
Cranmer 1999).
 
\subsection{Wind tilt angle as an eigenvalue and solution topology}

The critical point conditions for a specific streamline are, from
eq.~(\ref{eulerdiskwind})
 \begin{mathletters}
 \label{critpointconddisk}
 \begin{eqnarray}
 \label{critpointconddiska}
 W'\cri &=& {\alpha \over 1-\alpha} \; g\cri,\\
 \label{critpointconddiskb}
 E\cri &=& {1 \over \alpha^\alpha (1-\alpha)^{1-\alpha}} \;
{g\cri^{1-\alpha} \over f\cri},\\
 \label{critpointconddiskc}
 0&=&(1-\alpha)\;{g'\cri \over g\cri}-{f'\cri \over f\cri}.
 \end{eqnarray}
 \end{mathletters}

The tilted disk wind is essentially a 2-D phenomenon, and hence we
expect two eigenvalues of the Euler equation, with respect to $E$ and
$\lambda$. Finding the critical solution of maximum mass loss at a
given footpoint $r_0$ implies minimizing $E$ in
eq.~(\ref{critpointconddiskb}) with respect to the position of the
`critical' point $(\XX\cri,\lambda\cri)$. We show now that
$(\XX\cri,\lambda\cri)$ is a {\it saddle} point of $g^{1-\alpha}/f$.
We consider first the $\XX$ coordinate, and recall from the analysis
of the vertical disk wind that the maximum of $g^{1-\alpha}$ has
determined the eigenvalue $E\cri$. From
eq.~(\ref{critpointconddiskb}), the relevant function now is
$g^{1-\alpha}/f$. This means that the maximum of $g^{1-\alpha}/f$ with
respect to $\XX$ for a fixed $\lambda$ serves as a bottleneck of the
flow, i.e., the stringentest condition on the wind between the
photosphere and infinity. It, therefore, defines the maximum allowable
mass loss rate. Next, we analyze the mass loss rate along a streamline
by varying its tilt angle $\lambda$. To obtain the maximum mass loss
rate, we look for the minimum of $g^{1-\alpha}/f$ as a function of
$\lambda$. This particular $\lambda\cri$ plays the role of a second
eigenvalue of the Euler equation, besides $E\cri$. Note that due to
the dependency of $f$ on $r_0$, the wind tilt will change with $r_0$.
The eigenvalue $E\cri$ is thus given by
 \begin{equation}
 \label{minmax}
 E\cri = {1 \over \alpha^\alpha (1-\alpha)^{1-\alpha}} \;
\min\limits_\lambda \; \max\limits_\XX \; {g^{1-\alpha} \over f}.
 \end{equation} 
 \noindent This is the definition of a saddle point of
$g^{1-\alpha}/f$. Isocontours of this function are shown in
Fig.~\ref{lambdaeigenvalue} for the SHS disk. The existence of the
saddle point in $g^{1-\alpha}/f$ underlines the 2-D nature of disk
LDWs. Because the saddle point opens in the $\XX$-direction, the wind
escapes to large $\XX$.

Furthermore, the critical solution of maximum mass loss passes also
through a saddle point of the Euler function $P$ in the $(W',\XX)$
plane, in complete analogy with O~star winds. The regularity
condition, eq.~(\ref{critpointconddiskc}), determines the loci
$\XX\cri$ of these critical points, shown by the heavy lines in
Fig.~\ref{lambdaeigenvalue}. On the left branch of these curves, which
passes also through the saddle point of $g^{1-\alpha}/f$ (if the
latter exists), lie critical points of the saddle or X-type. Here,
$W'(\XX)$ can switch from a shallow (small $W'$) to a steep (large
$W'$) solution. On the other hand, the right branch of the regularity
curves, which passes through the minimum of $g^{1-\alpha}/f$, consists
of critical points of the focal type (Holzer 1977; Mihalas \& Mihalas
1984). They correspond to solutions which do not extend from the disk
photosphere to large radii, and are ignored in our discussion.

\begin{figure}
 \epsscale{.8}
 \plotone{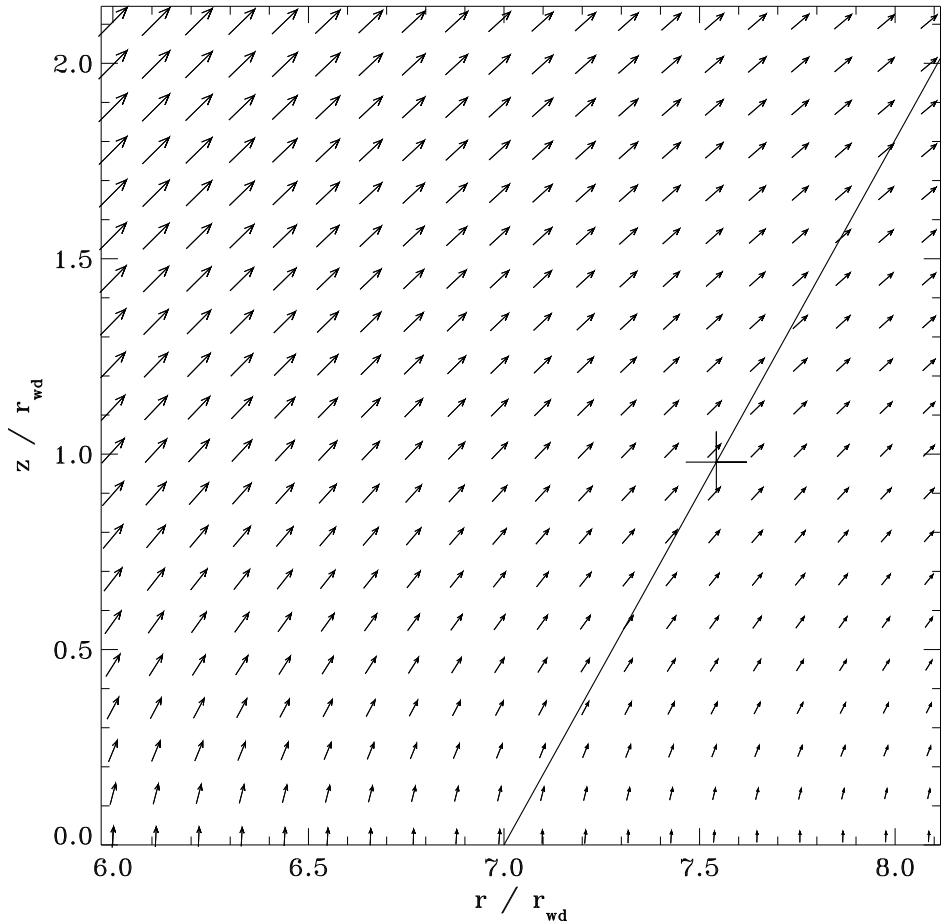}
 \caption{\label{fluxrayalign} Radiative flux vector above the SHS
disk ($\rd=30\rwd$). The straight line indicates a wind cone at
eigenvalue $\lambda=61\arcdeg$ (cf.~Table~1). The `$+$' sign marks the
critical point.}
 \end{figure}

Figure~\ref{fluxrayalign} shows a good overall alignment of the wind
ray of maximum mass loss rate with the radiative disk flux vector, at
least up to the critical point. This is (i) because the eigenvalue
$E\cri$ depends linearly on $f$, but only with a small power of
$1-\alpha$ on $g$, and (ii) because only $f$ has a maximum as function
of $\lambda$, whereas $g$ falls off monotonically.

\subsection{Inner and outer disk winds}

Up to this point we ignored the possibility of multiple saddle points
of $g^{1-\alpha}/f$. We now address this issue. As shown in
Fig.~\ref{lambdaeigenvalue}, for $r_0 \lea 4\rwd$ the function
$g^{1-\alpha}/f$ has only one saddle at a large height, e.g.,
$\XX\cri \simeq 4.4$ for $\alpha=2/3$. However, for $r_0 \gap
4\rwd$, a second saddle exists at smaller $\XX\cri$, which lies on
a different branch of the regularity curve. We name these two types of
saddles the {\it high} and {\it low} saddle, according to their height
$\XX\cri$ above the disk. The effective gravity `hill' separates
the two saddle points.

>From Fig.~\ref{lambdaeigenvalue}, the low saddle corresponds to a
larger mass loss rate than the high saddle. For $r_0 \gap 4\rwd$, the
solution of maximum mass loss is therefore determined by the low
saddle. For smaller $r_0$, however, only the high saddle exists, and
determines the wind solution then. These two cases define the {\it
outer} and {\it inner} disk wind, respectively. Clearly, the
assumption of straight streamlines is a severe one for the inner wind
with high-lying critical points.

The tilt angle of the outer wind is around $60\arcdeg$, namely
$\lambda\cri= 65\arcdeg$ at $r_0=4\rwd$, and $55\arcdeg$ at
$20\rwd$. This is largely independent of $\alpha$. For the inner wind,
the tilt is larger, $\lambda\cri= 80\arcdeg$ for $\alpha=2/3$ and
$70\arcdeg$ for $\alpha=1/2$.  Furthermore, the critical point,
$\XX\cri$, for the inner wind is much higher above the disk than the
critical point for the outer wind. As mentioned above, these critical
points fall on the opposite slopes of the effective gravity hill. For
the outer disk wind, the position $\XX\cri$ of the critical point is
moving closer to the wind sonic point with increasing $r_0$. The
reason for this is the larger gradient of the disk radiative flux in
the $\xx$-direction for larger $r_0$. As a result, the line force can
balance gravity at lower $\XX$.

Figure~\ref{accellaw} shows critical wind solutions $W'(\XX)$ above
the SHS disk, for different $r_0$. The decelerating solution branches,
$W'<0$, are discussed in Appendix~B. The critical point topology of
Fig.~\ref{accellaw} can be compared with that of the CAK stellar wind
flow in Fig.~\ref{caktopology}. (Note, that $W'$ has a slightly
different definition for the stellar and disk wind cases.) From
Fig.~\ref{accellaw}, we can also derive a condition for the {\it
existence} of a stationary, outer wind solution, further clarifying
the role of the effective gravity hill. The plus signs at the critical
points in the figure indicate where the Euler function $P>0$, i.e.,
where drag forces (gravity and inertia) overcome the driving forces
(line and centrifugal force), and correspondingly for the minus signs.
Hence, $\partial^2 P/\partial \XX\cri^2 <0$ at the low saddle, or,
using eq.~(\ref{critpointconddisk}), $(1-\alpha)\, g\cri''/g\cri <
f\cri''/f\cri$ (resp.~`$>$' at the high saddle). This means, the
maximum of $f$ must be sufficiently broad to allow for a stationary
solution with a low saddle. Note that the critical point for a
vertical wind above an infinite, isothermal disk, where $f=1$,
corresponds necessarily to a low saddle.

\begin{figure}
 \epsscale{1.05}
 \plotone{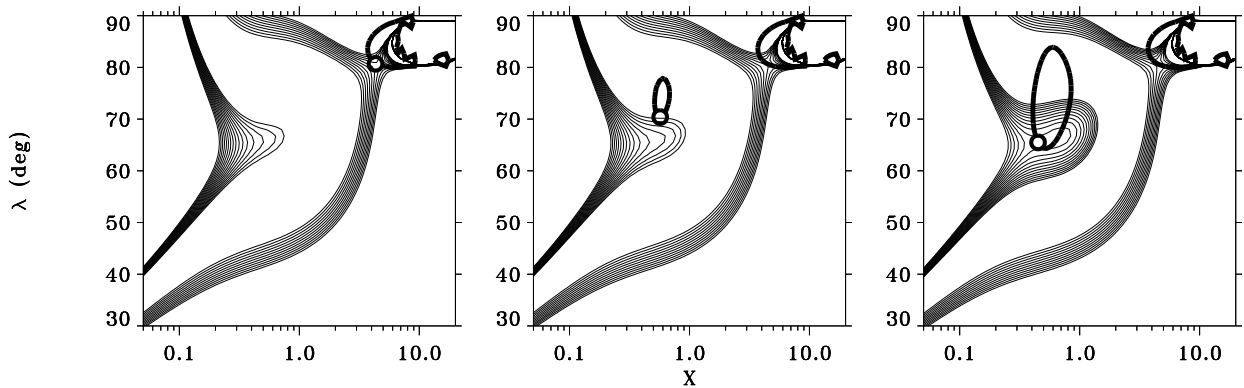}
 \caption{\label{switch80to60} Isocontours of $g^{1-\alpha}/f$,
normalized to $(2/3\sqrt{3})^{1-\alpha}$; from 0.855 to 0.885 in steps
of $2\times 10^{-3}$ at the low saddle; from 1.05 to 1.15 in steps of
$10^{-2}$ at the high saddle. At $r_0=4$, 4.03, and $4.15\rwd$ (left
to right panel). Heavy lines are solutions to the regularity
condition. Circles mark the critical wind solution of maximum mass
loss rate. Within the footpoint range from $r_0=4$ to $4.15\rwd$, the
wind switches from $\lambda\cri=80\arcdeg$ to $\lambda\cri=65\arcdeg$
via a growing regularity curve of ellipsoidal shape.}
 \end{figure}

In order to fully understand the geometry of disk LDWs, we consider
also the transition region between the inner and outer winds. As
discussed above, the low saddle does not exist below $r_0\lea
4\rwd$. Fig.~\ref{switch80to60} shows $g^{1-\alpha}/f$ in the
neighborhood of this footpoint radius. At $r_0=4\rwd$, only the high
saddle exists and determines the wind solution. At $r_0=4.03\rwd$, an
inner regularity curve of elliptical shape has formed, but not yet the
low saddle point of $g^{1-\alpha}/f$. The mass loss rate is maximal at
the smallest $\lambda$ along the curve, i.e., at its lower tip, which
determines the wind solution in this transition regime. Finally, by
$r_0=4.15\rwd$, a low saddle has formed at $\lambda\cri=65\arcdeg$.
The wind tilt stays at this value $\lambda\cri$, which corresponds to
the maximum mass loss rate. In total, the wind tilt switches
continuously from the high to the low saddle over a narrow range of
$0.1\rwd$ in the footpoint radius.

\subsection{Overall disk wind geometry}

Table~1 lists important parameters of the wind above SHS and Newtonian
disks, i.e., the tilt angle, $\lambda$, the normalized mass loss rate
from a disk annulus, $\dot m$, and the critical point location,
$\XX\cri$. The mass loss rate $\dot m$ is normalized to a vertical
wind above an isothermal disk. The shallow maxima of the function
$\FF_\xx$ in Fig.~\ref{ffunctionplot} are responsible for $\dot
m=O(1)$. Implications of these mass loss rates are discussed in
Paper~II. From the table, one finds the ray dispersion in the outer
wind, at intermediate footpoint radii from 4 to $10\rwd$,
 \begin{equation}
 \label{dlambdaoverdr}
 {d\lambda \over dr_0}\simeq -{1\arcdeg \over \rwd}.
 \end{equation}
 \noindent Further in or out the ray dispersion is even
smaller. Actually, since $d\lambda/dr_0$ enters the Euler equation
(\ref{eulerdiskwind}), the full wind problem can be solved only
iteratively. However, the dependence of the eigenvalues $E\cri$ and
$\lambda\cri$ on $d\lambda/dr_0$ is weak, and we assume throughout
that eq.~(\ref{dlambdaoverdr}) holds.

\begin{figure}
 \epsscale{1.}
 \plotone{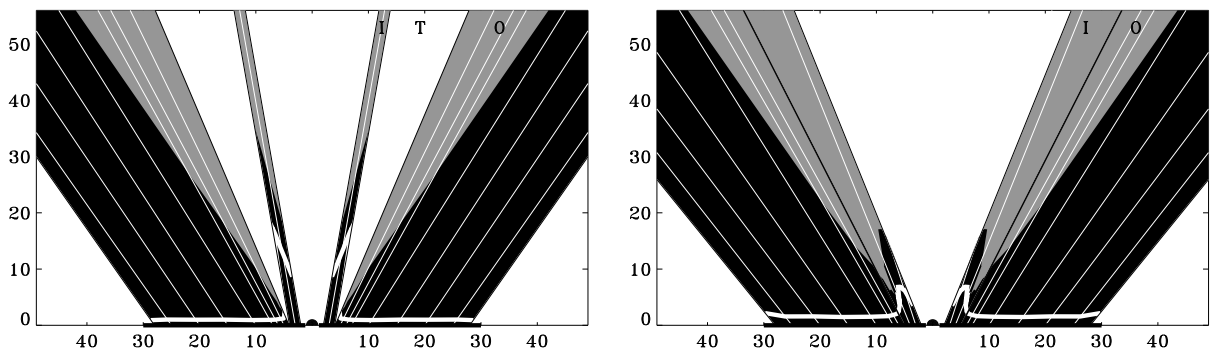} 
 \caption{\label{windsketch} Wind geometry above the SHS disk
($\rd=30\rwd$) according to Table~1, for $\alpha=2/3$ (left panel) and
1/2 (right panel). Black regions indicate the accelerating LDW, and
thin white lines show a few wind cones. Grey areas indicate
decelerating wind (Appendix B), for a ray dispersion $d\lambda/dr_0=
-1\arcdeg/\rwd$. `I' and `O' mark the inner and outer wind, `T' is the
transition region. Heavy, white lines are locations of flow critical
points. The innermost region of the disk, at $r_0<2\rwd$, is not
treated due to uncertainties in the radiation field.}
 \end{figure}

The overall geometry of the disk wind is shown in
Fig.~\ref{windsketch}. For $\alpha=2/3$, the critical points are at
$\xx\cri\sim 10-20 \rwd$ for the inner wind, then move towards the
disk photosphere and stay at $\xx\cri \simeq \rwd$, independent of
footpoint radius $r_0$ in the outer wind.  For $\alpha=1/2$, on the
other hand, the critical points lie somewhat higher for the outer
wind, at $\xx\cri \simeq 2\rwd$, but again independent of radius.
While the division between the inner and outer wind persists (namely
high-lying vs.~low-lying saddle, or critical points on opposing sides
of the gravity hill), the transition in $\lambda$ between the two
regions is smooth for $\alpha=1/2$, and the inner tilt reaches a
maximum of $\lambda=70\arcdeg$.

The innermost disk region, $1-2\rwd$, is left out in
Fig.~\ref{windsketch}. The details of the disk wind and its very
existence here are subject to great uncertainties in the radiation
field which depends on the properties of the transition layer and the
white dwarf itself. The outer boundary of the disk LDW is set by the
radius where the disk temperature falls below $10^4$~K and the wind
driving becomes inefficient, in analogy with stellar winds (Abbott
1982; Kudritzki~et~al.~1998). For the SHS disk with $L_{\rm d}=
10\Lsun$, this should happen around $30\rwd$.

\section{Discussion}
\label{discussion}

We compare here our theoretical model of LDWs from accretion disks in
CVs with those available in the literature, kinematical and dynamical
ones. We ignore the radial wind models, with the white dwarf being the
wind base, because they are in a clear contradiction with current
observations (e.g., review by Mauche \& Raymond 1997).  An alternative
source of gas is the disk itself.  Kinematical models which account
for this source of material subject to the line-driving force
successfully explained the observed bipolarity of the outflow, and
reproduced the inclination-dependent line profiles (Shlosman \&
Vitello 1993). Their weak point was the absence of a unique solution.
The 1-D dynamical models in a simplified disk radiation field revealed
some major differences between the stellar and disk winds, e.g., the
bi-polarity and the existence of a gravity hill (Vitello \& Shlosman
1988). 

More sophisticated 2-D kinematical models, supplemented with a 3-D
radiation transfer in Sobolev approximation, showed the importance of
rotation in shaping the lines (Vitello \& Shlosman 1993;
Shlosman~et~al.~1996). Finally, the 2-D hydrodynamical model of a disk
wind in a realistic radiation field and with the line-force
parameterized by the CAK approximation has addressed the issue of flow
streamlines and mass loss rates in the wind (PSD). Our comparison,
therefore, is focused on these models.
 
Vitello \& Shlosman (1993) set up a kinematical disk wind model
assuming straight flow lines in order to fit the C~{\sc iv} P~Cygni
line profiles of three CVs observed with the IUE.  The fit parameters
included the inner and outer terminating radius of the wind base, and
the corresponding tilt angles of the wind cone.  The best fit appeared
to be {\it indifferent} to the mass loss rate, within the range of
$10^{-1}$ to $10^{-2}$ of the accretion rate. In the present work,
which accounts for wind dynamics, we find lower mass loss rates more
justified and discuss various implications of these rates on the wind
models in Paper~II.  The tilt of the innermost wind cone in Vitello \&
Shlosman was rather steep, $\lambda= 80\arcdeg$, while at the outer
disk edge $\lambda= 25\arcdeg$. A similar work by
Knigge~et~al.~(1995), but using Monte Carlo radiation transfer in the
wind, gave similar results. In the present work, the tilt angle
$\lambda$ is calculated self-consistently from the Euler equation,
resulting in a similar inner tilt as found from kinematical models,
while the outer tilt differs by a factor of 2 between the two
approaches.

The most advanced numerical modeling of CV winds from the SHS disk so
far was performed by PSD, using the time-dependent Zeuss 2-D code. We
find a number of similarities between their and our results, but
differences exist as well. Our comparison with PSD is limited to their
models~2--5, i.e., without a central luminous star. These models agree
with ours on the overall wind geometry. This includes the streamline
shapes and the run of the wind opening angle with height. The
streamlines in PSD appear to form straight lines in the $(r,z)$ plane,
in striking similarity with the previous kinematical models. In
addition, the change in the wind opening angle with distance from the
rotation axis seems to be weak in PSD. The mass loss rates are
consistent between both models, and so are the wind optical depths,
which can approach unity even for very strong resonance lines
(Paper~II).

While PSD also find two markedly distinct flow regions, the inner and
outer, their inner wind, at $r_0\lea 4\rwd$, appears as the sole and
only outflow. The outer disk region, at radii $\gap 4\rwd$, exhibits a
time-dependent irregular flow, resulting in no mass loss. In contrast,
in our model, mass loss from the SHS disk is dominated by the inner
wind and the innermost part of the outer wind, as is discussed in
Paper~II.  Interestingly, our outer wind seems to be more robust than
the inner wind. For the latter, the balance of driving and drag forces
which leads to a high saddle on the far side of the gravity hill is
rather a delicate one. Setting, for example, the centrifugal force
{\it arbitrarily} to zero causes the high saddle solution to vanish,
whereas the low saddle remains almost unchanged.

PSD suggest that the irregular behavior of the outer flow is a
consequence of the different $\XX$-dependence of gravity and disk
flux, with the gravity preventing the wind from developing. We find
instead that at radii $r_0 \gap 4\rwd$, where a low saddle exists, the
fast increase in the projected disk flux, $\FF_\xx(\XX)$, results in a
sufficiently strong growth of the line force which drives the wind
past the gravity hill. For the inner wind regions, on the other hand,
where no lower saddle branch exists, the wind indeed must overcome the
gravity barrier without the appropriate radiation flux increase with
$\XX$.

Based on our analysis of the disk wind we may provide some insight
into the erratic behavior of the flow PSD observe at larger radii. It
is possible that mass overloading of the flow causes spatial and
temporal fluctuations of the streamline divergence $d\lambda/dr_0$,
wherefore the wind stalls on characteristic length scales of the
vertical gravity, $r_0$. We conjecture that fluctuations on such
scales may be self-amplifying, or, in other words, result in a locally
converging flow with $d\lambda/dr_0>0$. However, if such an
instability exists, it is expected to be most pronounced in the inner
disk regions. Namely, with increasing $r_0$, the flux increases faster
with $\XX$, which moves the critical point closer to the sonic
point. Mass overloading seems therefore less likely in the outer wind
regions, in contrast to the findings of PSD. A linear stability
analysis of our stationary wind model with respect to harmonic
perturbations of $d\lambda/dr_0$ would also be interesting, but is
beyond the scope of the present work.

Furthermore, we cannot confirm the dependence of $\lambda$ on the disk
luminosity as in the PSD model. We find that the eigenvalue
$\lambda\cri$ for each streamline is determined from the positions of
the saddle points of the function $g^{1-\alpha}/f$. Both $g$ and $f$
are independent of the disk luminosity, $f$ specifically so because it
is normalized to the flux at the streamline footpoint
(eq.~\ref{ffunctioneq}). Therefore, $\lambda\cri$ depends only on the
radial temperature stratification in the disk.

One important issue neglected in our modeling is the saturation of the
line force at some value when all the driving lines become optically
thin. If this thick-to-thin transition occurs before the flow reaches its
critical point, the wind solution is lost, since the drag forces
overcome the driving forces. However, this still leaves the
possibility that a more complicated wind dynamics is established,
where the decelerating flow at some larger radius starts again to
accelerate (i.e., jumps from a $W'_-$ to a $W'_+$ solution). We leave
this question open for future scrutiny, and note here that the mass
loss rates derived from the present eigenvalues $E$ are upper limits.

The present work is based on the CAK theory for stellar winds. Over
the years, questions have been raised concerning the physical meaning
of the CAK critical point (Thomas 1973; Lucy 1975; Cannon \& Thomas
1977; Abbott 1980; Owocki \& Rybicki 1986; Poe, Owocki, \&
Castor~1990). Most interesting for the present context is the
inclusion of higher order corrections to the diffuse line force in the
Sobolev approximation, which shift the critical point still closer to
the sonic point (Owocki \& Puls 1998; see also
Fig.~\ref{caktopology}).  This proximity of the sonic and critical
points may not be coincidental, and one can speculate whether or not
the {\it sonic} point determines the mass loss rate instead of the
critical point. However, we find for the disk wind model that the
sonic and critical points occasionally lie far apart, e.g., for a
vertical wind above an isothermal disk, or a tilted wind close to the
rotation axis (`inner wind').

These fundamental issues impair our understanding of LDWs from stars
and disks, and therefore must be addressed in the future.

\section{Summary}
\label{summary}

We discuss an analytical model for 2-D stationary winds from accretion
disks in cataclysmic variable stars. All parameters chosen are typical
for high-accretion rate disks in novalike CVs. We solve the Euler 
equation for
the wind, accounting for a realistic radiation field above the disk,
which drives the wind by means of radiation pressure in spectral
lines. Some key assumptions are that each helical streamline lies on a
straight cone; that the driving line force can be parameterized
according to CAK theory; and that the thermal gas pressure in the
supersonic wind can be neglected. Our results can be summarized as
follows.

The disk wind solutions are characterized by two eigenvalues, the mass
loss rate and the flow tilt angle, $\lambda\cri$, with the disk. The
additional eigenvalue $\lambda\cri$ for each streamline reflects the
2-D nature of the model. We find that the wind exhibits a clear
bi-conical geometry with a small ray dispersion. Specifically, two
regions, inner and outer, can be distinguished in the wind, launched
from within and outside $4\rwd$, respectively. The tilt angle for the
outer wind is $\lambda \sim 60\arcdeg$ with the disk. At these
angles, the wind flow and radiative disk flux are well aligned. For
the inner wind, the tilt angle is larger, up to $80\arcdeg$. We
emphasize that the disk wind tilt angle (i.e., the wind collimation)
depends solely upon the radial temperature stratification in the disk,
unless there is an additional degree of freedom such as central
luminosity associated with nuclear burning on the surface of the white
dwarf.

A major distinction between stellar and disk winds is the existence of
maxima in both the gravity and the disk flux along each streamline.
The latter flux maximum appears to be a crucial factor in allowing the
wind to pass over the gravity `hill'. The flux increase is more
pronounced further away from the rotation axis. As a result, the
critical point of the outer wind lies close to the disk photosphere
and to the sonic point. In fact, it lies upstream of the top of the
gravity hill, and this proximity of the critical and sonic points is
typical of LDWs from O~stars as well. On the other hand, for the inner
wind, the increase in radiation flux with height is smaller, and the
critical point lies far away from the sonic point, beyond the gravity
hill.

Comparing our analytical models with the 2-D numerical simulations of
Proga~et~al.~(1998), we find an overall good agreement in the
streamline shape, tilt angle, and mass loss rate, but our wind
baseline is wider.

\acknowledgments
We are grateful to Jon Bjorkman, Rolf Kudritzki, Chris Mauche, Norman
Murray, Stan Owocki, Joachim Puls, and Peter Vitello for numerous 
discussions on various aspects of line-driven winds. I.~S.~acknowledges 
hospitality of the IGPP/LLNL and its Director Charles Alcock, where this 
work was initiated. This work was supported in part by NASA grants 
NAG5-3841 and WKU-522762-98-06, and HST AR-07982.01-96A.

\appendix

\section{Line force due to gradients in the azimuthal velocity}

We estimate here the importance of azimuthal velocity terms for the
line force in $\xx$-direction. Assuming Keplerian rotation within the
disk, and angular momentum conservation above the disk, one has
 \begin{eqnarray}
 \label{vphigradients}
 {\dis\partial v_\phi/\partial z \over \dis\partial v_\xx/\partial
\xx} &=&-{\dis 1\over\dis\sqrt{2}} \;{\dis 1\over\dis \tala} \;{\dis
1\over\dis 1+\XX \cola}\; {\dis\sqrt{W}\over \;\dis W'},\nonumber\\
{\dis\partial v_\phi/\partial r \over \dis\partial v_\xx /\partial
\xx} &=&-{\dis 1\over\dis\sqrt{2}}\; {\dis 1-\XX\cola \over \dis
(1+\XX\cola)^2}\; {\dis\sqrt{W}\over \;\dis W'},\\ -{\dis v_\phi/r
\over \dis\partial v_\xx/\partial \xx} &=&- {\dis \sqrt{2} \over \dis
(1+\XX\cola)^2}\; {\dis\sqrt{W}\over \;\dis W'}\nonumber.
 \end{eqnarray}
 \noindent Here, the singularity of $\tan^{-1}\lambda$ at $\lambda=0$
is a result of neglecting the pressure terms in the Euler
equation. Note that $\partial v_\phi/\partial r$ changes sign at
$\XX=1/\cos\lambda$. From (\ref{vphigradients}), gradients in $v_\phi$
are comparable to gradients in $v_\xx$ when $\sqrt{W}/W' \sim 1$. The
main question is for their influence on the mass loss rate. Because,
in the CAK model, $\dot M$ is determined by the conditions at the
critical point, we calculate $\sqrt{W}/W'$ at the latter. We consider
first a vertical wind from an isothermal disk. Since $W'$ grows
monotonically up to and somewhat beyond the critical point (see
Fig.~\ref{accellaw}), and because $W=\int W' d\XX$, one has
$\sqrt{W\cri}/W'\cri<$ $(3/2)^{3/4} \sqrt{1-\alpha/ \alpha} \sim
1$. Here, $\XX\cri=1/\sqrt{2}$ and eq.~(\ref{eigenvalueverticalwinda})
were used. Alternatively, the critical points of the outer wind above
a non-isothermal disk typically lie close to the disk, hence
$g(\XX)\simeq \XX$. Using eq.~(\ref{critpointconddiska}),
$\sqrt{\XX\cri/ W'\cri}\sim \sqrt{1-\alpha/ \alpha} \sim 1$. Both disk
cases give, therefore, essentially the same result. We conclude that
$v_\phi$-terms can be important everywhere between the disk
photosphere and the critical point, and hence can modify $\dot M$.

To find their effect on $\dot M$, we include $v_\phi$-terms in the
evaluation of the line force, eq.~(\ref{lineforcegeneral}), in an
approximate manner. Only the disk regime is considered, in which case
the radiation intensity is roughly isotropic and the radiation flux
has a $z$-component only. The azimuthal part of the solid angle
integral in (\ref{lineforcegeneral}) is approximated by a 4-point
quadrature at angles $n\pi/2$ with ${\bf\hat r}$, where $n=0, 1, 2,
3$. This leads to a correction factor of the approximate form ${1\over
4} \bigl(2+|1+S|^\alpha+ |1-S|^\alpha\bigr)$ to the line force,
$EfW'^\alpha$. Here, $S$ is a linear combination of the expressions in
eq.~(\ref{vphigradients}), with coefficients $<1$ from angle
integration. In the disk regime, $S\cri\sim 1$ from
eq.~(\ref{vphigradients}). This coincides with the borderline between
an increase and a decrease in $\dot M$ due to inclusion of
$v_\phi$-terms, which lies at $S\cri= 1.25$ for $\alpha=1/2$ and at
$S\cri= 1.18$ for $\alpha=2/3$. A detailed, numerical calculation of
the above angle integral is required to decide, which of both cases
actually occurs. Since, however, $S\cri$ is close to unity, the
influence on the mass loss rate is limited to a 10\% effect. We,
therefore, neglect $v_\phi$-terms in calculating the line force.

\section{Disk wind deceleration}

\begin{figure}
 \epsscale{1.05}
 \plotone{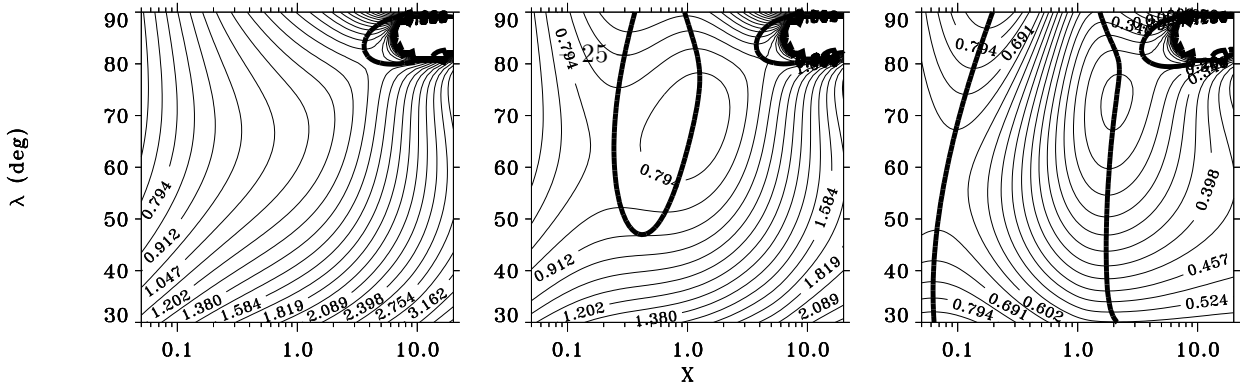}
 \caption{\label{lambdaeigenvalue} Isocontours of $g^{1-\alpha}/f$,
normalized to $(2/3\sqrt{3})^{1-\alpha}$, over the $(\XX,\lambda)$
plane. Footpoints of the wind are at $r_0=3$, 5, and $15\rwd$ (left to
right panel). The isocontours have logarithmic spacing. The
temperature stratification is that of the SHS disk ($\rd=30\rwd$,
$\alpha=2/3$). Heavy lines are solutions to the regularity condition
(\ref{critpointconddiskc}).}
 \end{figure}

In Figure~\ref{lambdaeigenvalue}, isocontours which cross through the
low saddle point {\it loop} into one another at some larger height,
$\XX_{\rm d}$. At $\XX>\XX_{\rm d}$, one has $E>E\cri$ from the
figure, i.e., the allowed maximum mass loss rate in this region is
{\it smaller} than that at the saddle. At these distances, inertia and
gravity overcome the line force plus centrifugal force, and the wind
decelerates, $W'<0$. As is shown in Paper~II, the wind speed always
exceeds the {\it local} escape speed at $\XX_{\rm d}$, which implies
that the decelerating wind reaches infinity at a positive speed.

\begin{figure}
 \epsscale{.8}
 \plotone{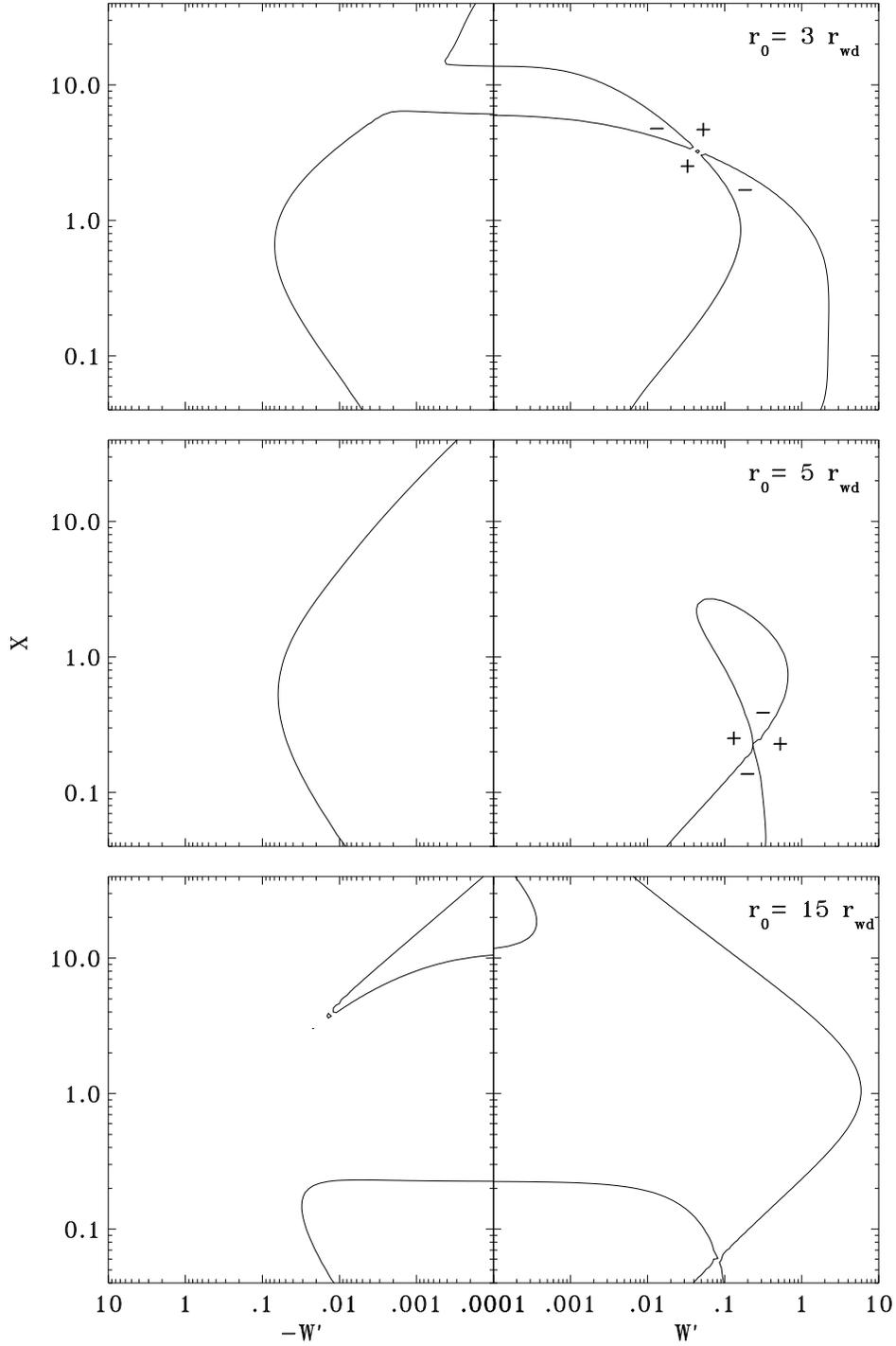}
 \caption{\label{accellaw} Critical wind solutions $W'(\XX)$ at zero
sound speed, above the SHS disk ($\rd=30\rwd$, $\alpha=2/3$). At
$r_0=3\rwd$: {\it high} saddle; at $r_0=5$ and $15\rwd$: {\it low}
saddle. The branches $W'<0$ are decelerating wind solutions, and are
discussed in Appendix~B.  `$+$' and `$-$' signs refer to the sign of
the Euler function $P$ in eq.~(\ref{eulerdiskwind}). For comparison
with the critical point topology of a spherically symmetric, stellar
wind, see Fig.~\ref{caktopology}.}
 \end{figure}

Due to the deceleration, the velocity law becomes non-monotonic, and
the line transfer is no longer purely local, because global couplings
occur between distant resonance locations. We neglect these couplings
here, and simply replace $W'^\alpha$ in the line force by
$|W'|^\alpha$. For a wind ray launched at $r_0=5\rwd$,
Fig.~\ref{accellaw} shows that a single, decelerating branch,
$W'_-<0$, accompanies the critical, accelerating solution $W'_+$ of
maximum mass loss rate. It is suggestive that at $\XX_{\rm d}$ the
solution curve jumps from the $W'_+$ to the $W'_-$ branch, and extends
thereupon to infinity.

The discontinuity in $W'$ introduces a {\it kink} in the velocity
law. Such kinks propagate at sound speed (Courant \& Friedrichs 1948;
actually, for LDWs, at some modified, radiative-acoustic speed, see
Abbott 1980 and Cranmer \& Owocki 1996) and are therefore inconsistent
with the assumption of stationarity. It seems plausible, however, that
the discontinuity in $W'$ is an artifact of the Sobolev approximation,
since the latter becomes invalid at small $dv/d\xx$, i.e., as
$W'\rightarrow 0$. An exact line transfer should instead give a smooth
transition from $W'_+$ to $W'_-$. We find indeed cases of `almost'
smooth transitions, where both $dW'_{+,-}/d\XX_{\rm d} \rightarrow
-\infty$, e.g., in the top panel of Fig.~\ref{accellaw}.

\clearpage

\begin{deluxetable}{cllllllllllll}
\tablecolumns{13}
\tablewidth{33pc}
%\tablewidth{0pc}
 \tablecaption{Tilt angle $\lambda\cri$, normalized mass loss rate
$\dot m\cri$, and position $\XX\cri$ of the critical point for
different disk wind models. Underlined numbers indicate the transition
from the inner to the outer wind.}
 \tablehead{
\colhead{} & \multicolumn{6}{c}{SHS Disk} & \multicolumn{6}{c}{Newtonian Disk}\\
\colhead{} & \multicolumn{3}{c}{$\alpha=2/3$}      & \multicolumn{3}{c}{$\alpha=1/2$} &
             \multicolumn{3}{c}{$\alpha=2/3$}      & \multicolumn{3}{c}{$\alpha=1/2$}\\
\colhead{$r_0/r_{\rm wd}$} &
\colhead{$\lambda\cri [\arcdeg]$} & \colhead{$\dot m\cri$} & \colhead{$\XX\cri$} & 
\colhead{$\lambda\cri$} & \colhead{$\dot m\cri$} & \colhead{$\XX\cri$} & 
\colhead{$\lambda\cri$} & \colhead{$\dot m\cri$} & \colhead{$\XX\cri$} & 
\colhead{$\lambda\cri$} & \colhead{$\dot m\cri$} & \colhead{$\XX\cri$} 
}
\startdata
 2 & 80 & 0.42 & 4.4        & 68 & 0.62 & 1.9       & 78 & 0.64 & 4.3       & 65 & 0.94 & 1.2        \nl
 3 & 80 & 0.60 & 4.4        & 72 & 1.02 & 2.3       & 78 & 0.90 & \ul{4.7}  & 65 & 1.45 & \ul{0.82}  \nl
 4 & 80 & 0.86 & \ul{4.4}   & 69 & 1.60 & \ul{1.7}  & 65 & 1.23 & 0.32      & 63 & 1.78 & 0.58       \nl
 5 & 64 & 1.37 & 0.26       & 63 & 2.23 & 0.52      & 64 & 1.32 & 0.26      & 62 & 2.10 & 0.43       \nl
 6 & 62 & 1.48 & 0.19       & 61 & 2.69 & 0.35      & 63 & 1.37 & 0.23      & 61 & 2.23 & 0.38       \nl
 7 & 61 & 1.60 & 0.16       & 60 & 3.08 & 0.27      & 63 & 1.37 & 0.21      & 61 & 2.37 & 0.34       \nl
10 & 58 & 1.82 & 0.11       & 58 & 3.84 & 0.18      & 62 & 1.48 & 0.17      & 60 & 2.69 & 0.28       \nl
15 & 57 & 2.10 & 0.08       & 55 & 5.41 & 0.12      & 61 & 1.54 & 0.15      & 58 & 3.08 & 0.23       \nl
20 & 56 & 2.26 & 0.06       & 53 & 6.57 & 0.10      & 60 & 1.60 & 0.13      & 58 & 3.31 & 0.2        \nl
25 & 55 & 2.45 & 0.05       & 52 & 8.16 & 0.08      & 60 & 1.67 & 0.13      & 57 & 3.31 & 0.3        \nl
28 & 55 & 2.59 & 0.04       & 51 & 8.16 & 0.10      & 58 & 1.35 & 0.2       & 57 & 2.44 & 0.2        \nl
\enddata
\end{deluxetable}

\clearpage

\end{document}